\documentclass[aps,prl, twocolumn, superscriptaddress, longbibliography]{revtex4-2}
\usepackage{amsmath}
\usepackage{amssymb}
\usepackage{graphicx}
\usepackage[caption=false, position=top, singlelinecheck=off, justification=raggedright]{subfig}
\usepackage{color}
\usepackage{pst-node}
\usepackage[unicode]{hyperref}
\hypersetup{unicode=true, colorlinks=true, linkcolor=blue, citecolor=blue}

\begin{document}

\title{Observation of a continuous time crystal}

\author{Phatthamon Kongkhambut}
\affiliation{Zentrum f\"ur Optische Quantentechnologien and Institut f\"ur Laser-Physik, Universit\"at Hamburg, 22761 Hamburg, Germany}

\author{Jim Skulte}
\affiliation{Zentrum f\"ur Optische Quantentechnologien and Institut f\"ur Laser-Physik, Universit\"at Hamburg, 22761 Hamburg, Germany}
\affiliation{The Hamburg Center for Ultrafast Imaging, Luruper Chaussee 149, 22761 Hamburg, Germany}

\author{Ludwig Mathey}
\affiliation{Zentrum f\"ur Optische Quantentechnologien and Institut f\"ur Laser-Physik, 
Universit\"at Hamburg, 22761 Hamburg, Germany}
\affiliation{The Hamburg Center for Ultrafast Imaging, Luruper Chaussee 149, 22761 Hamburg, Germany}

\author{Jayson G. Cosme}
\affiliation{National Institute of Physics, University of the Philippines, Diliman, Quezon City 1101, Philippines}

\author{Andreas Hemmerich}
\affiliation{Zentrum f\"ur Optische Quantentechnologien and Institut f\"ur Laser-Physik, Universit\"at Hamburg, 22761 Hamburg, Germany}
\affiliation{The Hamburg Center for Ultrafast Imaging, Luruper Chaussee 149, 22761 Hamburg, Germany}

\author{Hans Ke{\ss}ler}
\affiliation{Zentrum f\"ur Optische Quantentechnologien and Institut f\"ur Laser-Physik, 
Universit\"at Hamburg, 22761 Hamburg, Germany}

\begin{abstract}
Time crystals are classified as discrete or continuous depending on whether they spontaneously break discrete or continuous time translation symmetry. While discrete time crystals have been extensively studied in periodically driven systems since their recent discovery, the experimental realization of a continuous time crystal is still pending. Here, we report the observation of a limit cycle phase in a continuously pumped dissipative atom-cavity system, which is characterized by emergent oscillations in the intracavity photon number. We observe that the phase of this oscillation is random for different realizations, and hence this dynamical many-body state breaks continuous time translation symmetry spontaneously. The observed robustness of the limit cycles against temporal perturbations confirms the realization of a continuous time crystal.

\end{abstract}

\maketitle

\begin{figure}[h!]
	\centering
	\includegraphics[width= 1\columnwidth]{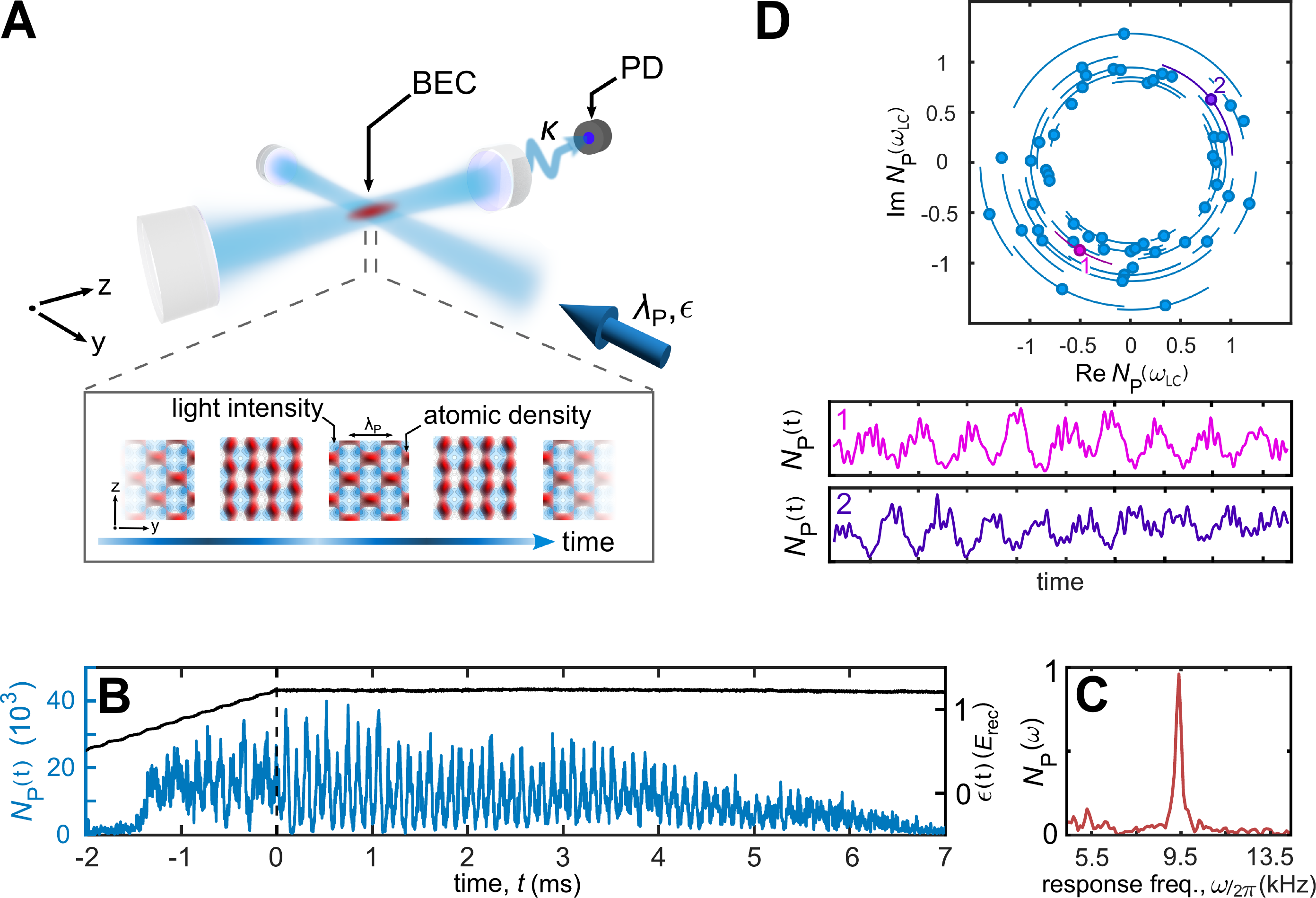}
	\label{fig:1} 

\caption{{\bf Continuous time crystal in an atom-cavity system.} ({\bf  A}) Schematic drawing of the atom-cavity system pumped transversely with an optical pump lattice, blue detuned with respect to an atomic transition. The inset in the bottom shows the photon field (blue) and the atomic density (red) of the limit cycle dynamics, based on simulations. The blue color shading of the time axis indicates the intracavity photon number. ({\bf B})  Single experimental realization of the limit cycle phase for $\delta_\mathrm{eff}/2\pi=-3.8$~kHz and $\epsilon_\mathrm{f}=1.25~E_\mathrm{rec}$. The vertical dashed black line indicates the start of the 10~ms holding time, wherein the pump strength is held constant. Black line: time trace of the pump strength $\epsilon$. Blue line: time evolution of the intracavity photon number $N_{\mathrm{P}}(t)$. ({\bf C}) Normalized and rescaled single-sided amplitude spectrum of $N_\mathrm{P}$ calculated from the data shown in B. ({\bf D}) Distribution of the time phase in the limit cycle phase for $\delta_\mathrm{eff}/2\pi=-5.0$~kHz and $\epsilon_\mathrm{f}=1.25~E_\mathrm{rec}$. The error bars represent the phase uncertainty within our discrete Fourier transform resolution of $100$ Hz. The uncertainty with regard to the radial dimension, i.e. the amplitude uncertainty, however, is negligibly small. For clarity, we remove the errors bars, around 30\%, which are overlapping. The two panels in the bottom show the evolution of the intracavity photon number for two specific experimental realizations, marked with 1,2 the upper panel, which have a time phase difference of almost $\pi$.}
\end{figure}

Time crystals are dynamical many-body states that break time translation symmetry in a spontaneous and robust manner \cite{Wilczek2012, Shapere2012}. The original quantum time crystal envisaged by Frank Wilczek involves a closed many-body system with all-to-all coupling that breaks continuous time translation symmetry by exhibiting oscillatory dynamics in its lowest energy equilibrium state even though the underlying Hamiltonian is time-independent \cite{Wilczek2012}. This would in fact constitute a startling state of matter in motion, fundamentally protected from bringing this motion to a standstill by energy removal. However, a series of no-go theorems have shown that nature prohibits the realization of such time crystals in isolated systems \cite{Noz2013, Bruno2013, Watanabe2015}. The search for time crystals was thus extended to include equilibrium scenarios in periodically driven closed systems \cite{Else20, Sacha2020, Khemani2019}. This has led to realizations of discrete time crystals, which break the discrete time translation symmetry imposed by the external drive \cite{Zhang2017, Choi2017, Rovny2018, Smits2018, Autti2018, Sullivan20, Monroe2021, Randall2021, Mi2021}. In such discrete time crystals, during a short initial phase, the drive slightly excites the system, until the system decouples from the drive, such that further energy or entropy flow is terminated. The system develops a subharmonic response, i.e., an intrinsic oscillation at a frequency slower than that of the drive. Initially, it was argued that dissipation, and hence the use of open systems, must be carefully avoided, until so called dissipative discrete time crystals were theoretically predicted \cite{Gong2018} and experimentally realized \cite{Kessler2021,Kongkhambut2021,Taheri22}. As shown in a number of theoretical works \cite{Iemini2018, Buca2019, Kessler2019}, the use of open systems comes with the surprising consequence that continuous instead of periodic driving suffices to induce time crystal dynamics. These continuous time crystals realize the spirit of the original proposal more closely than discrete time crystals and circumvent the no-go theorems via their open character.

Here, we report the observation of a continuous time crystal (CTC) in the form of a limit cycle phase in a continuously pumped dissipative atom-cavity system (cf. Fig.~1A). In classical nonlinear dynamics, the term limit cycle, coined by Poincar{\'e} in a mathematical context \cite{Poi:80}, denotes a closed phase space trajectory, asymptotically approached by at least one neighboring trajectory. While limit cycles are well-established in classical nonlinear physics \cite{Strogatz}, there are two essential conditions for limit cycles in open quantum systems to form a CTC. Firstly, the formation of the limit cycle must be associated with spontaneous breaking of continuous time translation symmetry. That is, the relative time phase of the oscillations for repeated realizations takes random values between 0 and 2$\pi$. Secondly, the limit cycle phase is robust against temporal perturbations of technical or fundamental character, such as quantum noise and, for open systems, fluctuations associated with dissipation. The characteristic signature of the CTC presented here is a persistent oscillation of the intracavity intensity and atomic density (Fig.~1B,C), which complies with the robustness and spontaneous symmetry breaking criteria (Fig.~1D). 

Our experimental setup consists of a Bose-Einstein condensate (BEC) of $N_\mathrm{a} \approx 5\times10^4$ $^{87}$Rb atoms inside a high-finesse optical cavity. The system is transversely pumped with a standing wave field with a wavelength $\lambda_\mathrm{P} = 792.55\,$nm (Fig.~1A). This wavelength is blue detuned with respect to relevant atomic D$_1$ transition of $^{87}$Rb at a wavelength of $794.98\,$nm. The cavity operates in the recoil resolved regime \cite{Kessler2014}, i.e. its field decay rate $\kappa = 2\pi\times3.4$~kHz is smaller than the recoil frequency $\omega_\mathrm{rec}=2\pi \times 3.7\,$kHz. The cavity resonance frequency $\omega_\mathrm{c}$ is shifted due the the refractive index of the BEC by an amount of $\delta_- = N_\mathrm{a} U_0/2$, where $U_0=2\pi\times1.3$~Hz is the maximal light shift per intracavity photon. We define the effective detuning $\delta_{\mathrm{eff}}\equiv \delta_{c}-\delta_-$ where $\delta_c \equiv \omega_\mathrm{p} - \omega_\mathrm{c}$ is the detuning between the pump field frequency $\omega_\mathrm{p}$ and the resonance frequency of the empty cavity $\omega_\mathrm{c}$.

To determine the regime of the CTC, we measure the time dependence of the intracavity photon number $N_\mathrm{P}(t)$ that emerges in the protocol given below. We display $N_\mathrm{P}(t)$ in Fig.~2A, and two derived quantities, the crystalline fraction $\Xi$, and the limit-cycle frequency $\omega_\mathrm{LC}$ in Figs.~2B and C, respectively. In our protocol, the intracavity photon number $N_\mathrm{P}(t)$ is recorded as we linearly ramp the pump strength $\epsilon$ from $0$ to $3.5\,E_\mathrm{rec}$ within 10~ms, while keeping $\delta_\mathrm{eff}$ fixed. Initially, for weak pump intensities, the BEC phase is stable and $N_\mathrm{P}$ is zero. Above a critical value of $\epsilon$, the BEC becomes unstable towards the formation of a self-organized superradiant phase heralded by a nonzero $N_\mathrm{P}$. This represents a many-body state as the cavity photons mediate a retarded infinite-range interaction between the atoms. While this superradiant phase transition has been intensively studied for a red-detuned pump \cite{Domokos2002, Black2003, Baumann2010, Klinder2015}, it has only been realized recently for a blue-detuned pump following its theoretical prediction \cite{Zupancic2019, Piazza2015}. For blue detuning, the atoms are low-field seeking and they localize at the intensity minima of the light field. Nevertheless, the atoms can still self-organize into the superradiant phase as evident from the large blue areas in Fig.~2A. However, the self-organized superradiant phase may become unstable for higher pump strengths, as it costs energy for the atoms to localize away from the nodes of the pump lattice. This behavior leads to the disappearance of the self-organized phase for higher pump strengths \cite{Zupancic2019}. Fig.~S1 in \cite{SM} shows a phase diagram for a larger range of $\epsilon$, demonstrating the disappearance of the self-organization for strong pumping. In the recoil-resolved regime, due to the retarded character of the cavity-mediated interaction, we additionally observe the emergence of a novel dynamical phase or a limit cycle phase characterized by self-sustained oscillations of $N_\mathrm{P}$ as the atoms cycle through different density wave patterns \cite{Piazza2015, Kessler2020}. The resolution of the experimental imaging system is insufficient to observe the real space density of the cloud, instead Fig.~S3 in the Supplemental Materials shows simulations of the evolution of the single-particle density using a mean-field model. Physically, the limit cycles can be understood as a competition between opposing energy contributions, one coming from the pump lattice potential and another from the cavity-induced all-to-all interaction between the atoms \cite{Piazza2015}. In the superradiant phase, the cavity-induced interaction energy dominates and the atoms localize at the antinodes. In the limit cycle phase for sufficiently strong pump intensities, localization of low-field seeking atoms at the antinodes becomes energetically costly, resulting in a decrease in the density modulations and $N_\mathrm{P}$ as the system attempts to go back to the normal homogeneous phase. However, this is unstable towards self-organization since the chosen pump strength already exceeds the critical value and thus, the cycle starts anew. The regime of recoil-resolution of the cavity, where the dynamics of the atomic density and the light field evolve with similar time scales, has turned out to be the key ingredient to realize the limit cycle phase. This can be understood by the fact, that the delayed dynamics of the cavity field, with respect to the atomic density, leads to cavity cooling, which in contrast to broadband cavity setups, restricts the atoms to occupy only a small number of momentum modes. This prevents the system from heating up and entering chaotic dynamics. In Fig.~2A, we observe the limit cycle phase in the region enclosed by the yellow dashed lines. To further highlight the dynamical nature of this phase, we show a typical single-shot realization in Fig.~1B and C.

\begin{figure*}[!th]
	\centering
	\includegraphics[width= 1.6\columnwidth]{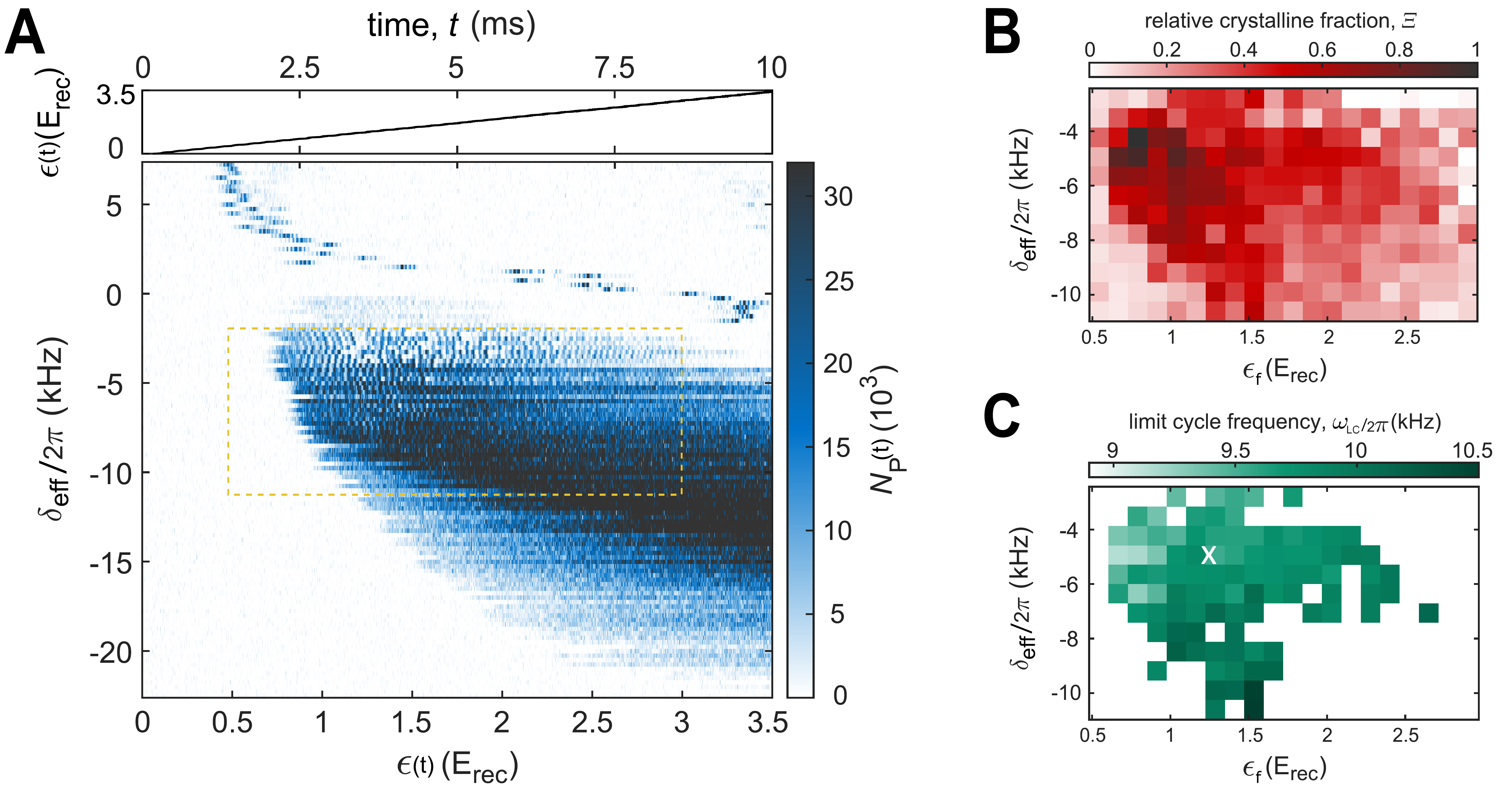}
	\caption{{\bf Determining the time-crystalline regime.} ({\bf  A}) Top panel: Pump strength protocol. Bottom panel: The corresponding intracavity photon number $N_\mathrm{P}$ as a function of $\delta_\mathrm{eff}$ and $\epsilon$. The area enclosed by the yellow dashed lines marks the parameter space spanned in B and C. ({\bf B}) Relative crystalline fraction $\Xi$ and ({\bf C}) limit cycle frequency $\omega_\mathrm{LC}$ plotted versus $\delta_\mathrm{eff}$ and $\epsilon_\mathrm{f}$. To obtain B and C, for fixed $\delta_\mathrm{eff}$, the pump strength is ramped to its final value $\epsilon_\mathrm{f}$, and subsequently held constant for 10 ms. The relative crystalline fraction $\Xi$ and the corresponding value of $\omega_\mathrm{LC}$ to identify the time-crystalline state. The parameter space is divided into 20 $\times$ 24 plaquettes and averages over 5 to 10 experimental implementations are produced. The white cross indicates the parameter values $\delta_\mathrm{eff}/2\pi=-5.0\,$kHz and  $\epsilon_\mathrm{f}=1.25\,E_\mathrm{rec}$. The white area in C corresponds to data with $\Xi$ below $1/e$.}
	\label{fig:2} 
\end{figure*}

Next, we quantitatively identify the area in the parameter space, spanned by the pump strength $\epsilon$ and the effective detuning $\delta_\mathrm{eff}$, where limit cycles can be observed. For fixed $\delta_\mathrm{eff}$, we linearly ramp $\epsilon$ to the desired final value $\epsilon_\mathrm{f}$, using the same slope as for the measurement presented in Fig.~2A, and hold $\epsilon$ constant for $10$~ms. The protocol is depicted by the black curve in Fig.~1B. We show in Fig.~1C an example of the normalized and rescaled single-sided amplitude spectrum $N_\mathrm{P}(\omega) = \overline{N}_\mathrm{P}(\omega)/ \overline{N}_\mathrm{P,max}(\omega_\mathrm{LC})$ obtained from $N_\mathrm{P}(t)$ within the holding time window $[0,10]\,\mathrm{ms}$ in Fig.~1B. $\overline{N}_\mathrm{P}(\omega)$ is the normalized single-sided amplitude spectrum and $\overline{N}_\mathrm{P,max}(\omega_\mathrm{LC})$ is the maximum value of the measured limit cycle amplitude. In the case of pronounced limit cycle dynamics as in Fig.~1C, the single-sided amplitude spectrum shows a distinct peak, with a width associated with the limit cycle lifetime of several milliseconds. The narrowest peaks observed exhibit a $e^{-2}$ width $\Delta \omega  \approx 2\pi \times 1.4\,$kHz: The limit cycle frequency $\omega_\mathrm{LC}$, plotted in Fig.~2C, is defined as the frequency of the dominant peak in the single-sided amplitude spectrum within the frequency interval $\Delta_\mathrm{LC} = [3.5,15.5]\times2\pi~$kHz, chosen much larger than $\delta_\mathrm{LC} \in [\omega_\mathrm{LC} - \Delta \omega/2,\omega_\mathrm{LC} + \Delta \omega/2]$. The oscillation frequency of a CTC is not necessarily fixed and robustness refers to the persistence of the CTC in the thermodynamic limit and for a wide range of system parameters \cite{Iemini2018} (Finite-size effects are discussed in the Supplementary Materials). We calculate a common measure for time crystallinity, the crystalline fraction $\Xi'$ \cite{Choi2017, Rovny2018}, as the ratio between the area under the single-sided amplitude spectrum within $\delta_\mathrm{LC}$ and the total area within $\Delta_\mathrm{LC}$. That is, $\Xi' \equiv \sum_{\omega \in \delta_\mathrm{LC}} N_\mathrm{P}(\omega)/\sum_{\omega \in \Delta_\mathrm{LC}} N_\mathrm{P}(\omega)$. The relative crystalline fraction $\Xi$ shown in Fig.~2B is normalized to the maximum crystalline fraction measured in the parameter space explored in this work. Due to the finite lifetime of the BEC, it is difficult to access the long-time behavior of the system, which makes it experimentally challenging to distinguish between the areas of stable limit cycle, chaos, and possible transient phases. Hence, we define a cut-off or threshold value for the relative crystalline fraction, $\Xi_\mathrm{cut}=1/e$, to identify regions with observable limit cycle dynamics. In Fig.~2C, the frequency response of the limit cycle phase is only shown if its relative crystalline fraction is higher than the cut-off value, i.e., $\Xi > \Xi_\mathrm{cut}$. The experimental lifetime of our time crystal is limited by atom loss. Furthermore, the short-range contact interaction, due to collisions between the atoms, leads to dephasing of the system and, hence, melting of the time crystal. Simulations including contact interactions and phenomenological atom loss can be found in the Supplementary Materials.

\begin{figure*}[h!t]
	\centering
	\includegraphics[width= 1.8\columnwidth]{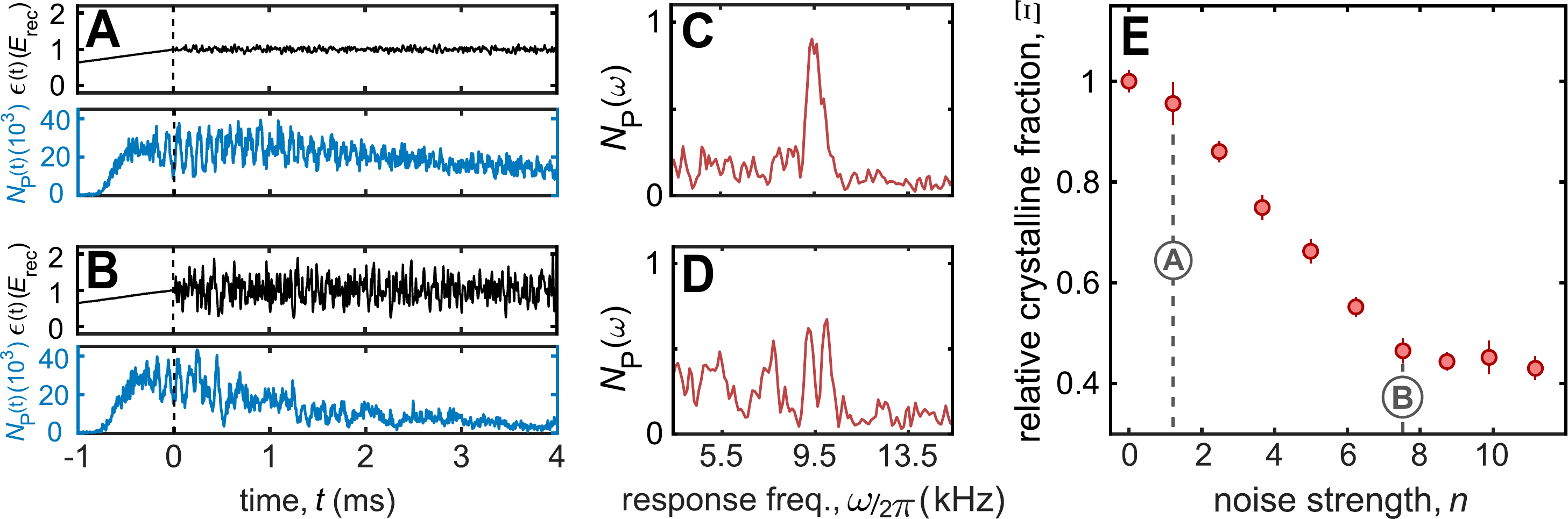}
	\caption{{\bf Robustness against temporal perturbations.} ({\bf A-B}) Single experimental runs for noise strengths indicated in (E). Top panels: time traces of the pump strength $\epsilon$. Bottom panel: corresponding dynamics of $N_\mathrm{P}$. ({\bf C-D}) Single-sided amplitude spectra of (A) and (B), respectively. ({\bf E}) Relative crystalline fraction for varying noise strength $n$ and fixed $\delta_\mathrm{eff}/2\pi=-5.0$~kHz and $\epsilon_\mathrm{f}=1.25~E_\mathrm{rec}$.}
	\label{fig:3} 
\end{figure*}
	
The spontaneous symmetry breaking of a many-body system indicates a phase transition. Here, we demonstrate strong evidence that the limit cycle phase emerges through spontaneous breaking of continuous time translation symmetry and thus, it is a CTC. We repeat the experimental protocol used in Fig.~1D for more than 1500 times with fixed $\delta_\mathrm{eff}/2\pi=-5.0\,$kHz and  $\epsilon_\mathrm{f}=1.25\,E_\mathrm{rec}$. These parameter values are indicated in Fig.~2C by a white cross. Due to technical instabilities, the number of the atoms in the BEC $N_\mathrm{a}$ fluctuates by 5$\%$. This leads to a fluctuating value of $\delta_\mathrm{eff}$ and hence of $\omega_\mathrm{LC}$. Pictorially, this can be understood by observing that fluctuations in $N_\mathrm{a}$ effectively shift the CTC regime in Fig.~2C either up or down. For the parameter values indicated by a white cross in Fig.~2C, the median of $\omega_\mathrm{LC}$ is $\overline{\omega}_\mathrm{LC} = 2\pi \times 9.69$ kHz. Our discrete Fourier transform resolution, set by the 10 ms time window, is $100$~Hz. Thus, we only consider experimental runs, which yielded response frequencies of $\omega_\mathrm{LC} = \overline{\omega}_\mathrm{LC}\pm\, 2\pi \times (50\,\mathrm{Hz})$. For each single-shot measurement, we obtain the time phase defined as the principal argument $\mathrm{arg}(N_\mathrm{P}(\omega_\mathrm{LC}))$ of the Fourier transformed intracavity photon number $N_\mathrm{P}(\omega_\mathrm{LC})$ evaluated at the limit cycle frequency $\omega_\mathrm{LC}$. In Fig.~1D, we present the distribution of the observed time phases, which randomly covers the interval $[0,2\pi)$. This corroborates the spontaneous breaking of continuous time translation symmetry in the limit cycle phase. In the bottom of Fig.~1D we present two specific experimental realizations, which having a time phase difference of almost $\pi$. Simulations representing the BEC as a coherent state show a range of the response frequency distribution of $300$~Hz. Since we post-select our data far below this limit, the origin of the spread over $2\pi$ in the time phase distribution is not due to technical noises but rather due to quantum fluctuations. In the Supplementary Materials, we show a more detailed theoretical analysis to support this argument. Note that the error bars along the angular direction in Fig.~1D represent the phase uncertainty within $100$~Hz of our Fourier limit. The average phase uncertainty is around $0.25\pi$. The uncertainty in the radial direction corresponding to the oscillation amplitude is, however, negligible. Moreover, we remove $30\%$ of the error bars for clarity in Fig.~1D.

Finally, we demonstrate the robustness of the limit cycle phase against temporal perturbations, which is a defining feature of time crystals. We introduce white noise onto the pump signal with a bandwidth of $50$~kHz. The noise strength is quantified by \newline $n \equiv \sum_{\omega=0}^{2\pi \times 50~\mathrm{kHz}} |\mathcal{A}_\mathrm{noisy}(\omega)|/\sum_{\omega=0}^{2\pi \times 50~\mathrm{kHz}} |\mathcal{A}_\mathrm{clean}(\omega)| -1$,  where $\mathcal{A}_\mathrm{noisy}$ ($\mathcal{A}_\mathrm{clean}$) is the single-sided amplitude spectrum of the pump in the presence (absence) of white noise. We choose the parameters $\delta_\mathrm{eff}/2\pi=-5.0$~kHz and $\epsilon_\mathrm{f}=1.25\,E_\mathrm{rec}$ in the center of the stable limit cycle region, indicated by the white cross in Fig.~2C, and add white noise with varying strengths. In the upper panels of Figs.~3A and 3B, single-shot realizations of the noisy pump signal are shown for weak and strong noise, respectively. The corresponding dynamics of $N_\mathrm{P}$ is shown in the bottom panels of the respective plots. In Fig.~3E, we show how increasing the noise strength can `melt' the CTC as inferred by the decreasing relative crystalline fractions calculated from single-sided amplitude spectra, similar to those shown in Figs.~3C and 3D. Note that the system takes time to react to the noise, such that a few oscillations can always be observed before decay sets in. This leads to an offset of 0.4 in the crystalline fraction even for very strong noise. Nevertheless, we find that the limit cycle phase indeed exhibits robust oscillatory behavior over a wide range of the noise strength. This, together with the observation of spontaneous breaking of a continuous time translation symmetry, suggests that the observed limit cycle phase is a CTC.

In conclusion, we have experimentally demonstrated a continuous time crystal, and provided a theoretical understanding. This class of dynamical many-body states expands the concepts of long-range order and spontaneous symmetry breaking into the time domain, and is therefore of fundamental interest. This result, and the exquisite precision and control achieved with our atom-cavity platform, paves the way towards a broad and comprehensive study of dynamical many-body states of bosonic or fermionic quantum matter in the strongly correlated regime. For example, an increased atom-photon coupling could generate a new class of time crystals associated with symmetry broken periodic entanglement. Furthermore, technological applications, e.g. towards time metrology, can be envisioned.

\section*{Acknowledgments}

H.K. thanks J. Klinder and C. Georges for helpful discussions and their support. J.G.C. thanks R. J. L. Tuquero for valuable insights and discussions. A.H. acknowledges useful discussions with C. Zimmermann and J. Marino.
\textbf{Funding:} This work is funded by the Deutsche Forschungsgemeinschaft (DFG, German Research Foundation) through grant DFG-KE$2481/1-1$. P.K., J.S., L.M. and A.H. acknowledge the DFG for funding through SFB-925 -- project 170620586, and the Cluster of Excellence “Advanced Imaging of Matter” (EXC 2056) -- project No. 390715994. J.S. acknowledges support from the German Academic Scholarship Foundation.
\textbf{Author contributions:}
P.K. and H.K. performed the experiments and data analysis. The simulations were performed by J.S. and J.G.C., supported by L.M.. The project was designed and supervised by H.K. and A.H.. All authors contributed to the discussion and interpretation of the results, as well as, to writing the manuscript.

\end{document}


\title{Supplemental Material for\\Observation of a continuous time crystal}

\author{Phatthamon Kongkhambut}
\affiliation{Zentrum f\"ur Optische Quantentechnologien and Institut f\"ur Laser-Physik, Universit\"at Hamburg, 22761 Hamburg, Germany}

\author{Jim Skulte}
\affiliation{Zentrum f\"ur Optische Quantentechnologien and Institut f\"ur Laser-Physik, Universit\"at Hamburg, 22761 Hamburg, Germany}
\affiliation{The Hamburg Center for Ultrafast Imaging, Luruper Chaussee 149, 22761 Hamburg, Germany}

\author{Ludwig Mathey}
\affiliation{Zentrum f\"ur Optische Quantentechnologien and Institut f\"ur Laser-Physik, Universit\"at Hamburg, 22761 Hamburg, Germany}
\affiliation{The Hamburg Center for Ultrafast Imaging, Luruper Chaussee 149, 22761 Hamburg, Germany}

\author{Jayson G. Cosme}
\affiliation{National Institute of Physics, University of the Philippines, Diliman, Quezon City 1101, Philippines}

\author{Andreas Hemmerich}
\affiliation{Zentrum f\"ur Optische Quantentechnologien and Institut f\"ur Laser-Physik, Universit\"at Hamburg, 22761 Hamburg, Germany}
\affiliation{The Hamburg Center for Ultrafast Imaging, Luruper Chaussee 149, 22761 Hamburg, Germany}

\author{Hans Ke{\ss}ler}
\affiliation{Zentrum f\"ur Optische Quantentechnologien and Institut f\"ur Laser-Physik, Universit\"at Hamburg, 22761 Hamburg, Germany}

\maketitle
 
\section*{Experimental details}
The experimental setup, as sketched in Fig.~1(A) in the main text, is comprised of a magnetically trapped BEC of $N_\mathrm{a} = 5\times 10^4$  $^{87}$Rb atoms, dispersively coupled to a narrowband high-finesse optical cavity. The trap creates a harmonic potential with trap frequencies $\omega = 2\pi \times (119.0,102.7,24.7)$~Hz. The cavity field has a decay rate of $\kappa=~2\pi \times3.4\,$kHz, which almost equals the recoil frequency $\omega_\mathrm{rec} = E_\mathrm{rec}/\hbar  =~2\pi \times3.7\,$kHz for pump wavelength of $\lambda_\mathrm{P} = 792.55\,$nm. The pump laser is blue detuned with respect to the relevant atomic transition of $^{87}$Rb at 794.98~nm. The maximum light shift per atom is $U_0 = 2\pi \times 1.3~\mathrm{Hz}$. A typical experimental sequence starts by preparing the BEC and linearly increasing the pump strength $\epsilon$ to its desired value $\epsilon_\mathrm{f}$ and subsequently holding it constant for $10$~ms.

\section*{Phase diagram for large pump strength range}

In Fig.~S1A we present a phase diagram, similar to the one shown in Fig.~2A in the main text, but for larger pump strength range. The experimental protocol is the same as for Fig.~2A but the ramp time is increased to $20$~ms. For strong pumping the system does not favor anymore the self-organization, since the cost of localizing the atoms at the nodes of the potential exceeds the decrease of energy due to the cavity-mediated coupling. In Fig.~S1B the phase difference between the pump and cavity field $\phi$ is plotted against  $\delta_\mathrm{eff}$ and $\epsilon$. In the self-organized phase, $N_\mathrm{P}$ is finite and $\phi$ locks to either 0 or $\pi$ and stay constant. In Fig.~S1C, we present the amplitude of the Fourier spectrum calculated from the photon number data. The limit cycle region can be identified by a peak in the frequency response around $10$~kHz.

\begin{figure}[h]
	\centering
	\includegraphics[width= 0.95\columnwidth]{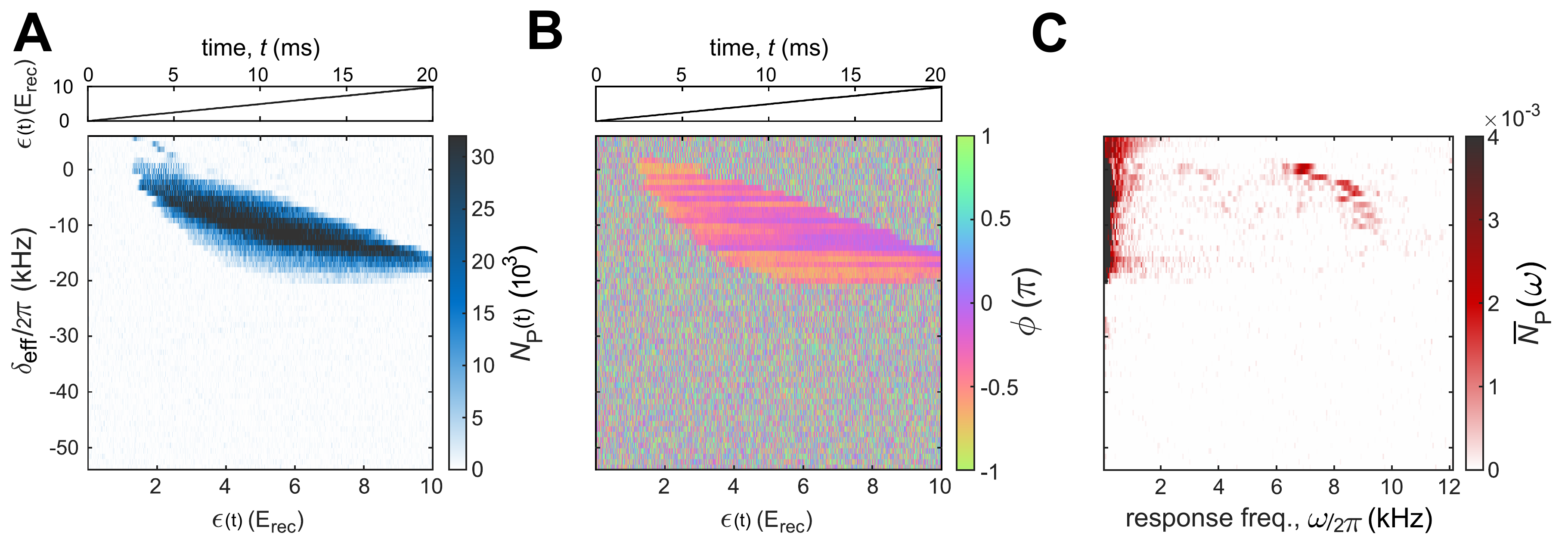}
	\label{sfig:1} 
\caption{{\bf Phase diagrams for large pump strength range.} ({\bf A}) Top panel: Pump strength protocol. Bottom panel: The corresponding intracavity photon number $N_\mathrm{P}$, as a function of the effective detuning $\delta_\mathrm{eff}$ and pump strength $\epsilon$ at a pump wavelength of $\lambda_\mathrm{P} = 792.55\,$nm. The corresponding light shift per photon is $U_0 = 2\pi \times 1.3~\mathrm{Hz}$. ({\bf B}) Top panel: Pump strength protocol. Bottom panel: The phase difference between the pump and intracavity field $\phi$, as a function of the effective detuning $\delta_\mathrm{eff}$ and pump strength $\epsilon$. Note, due to technical instabilities of the phase reference, we observe a drift of the phase signal of the cavity field of about $0.02\pi$ per ms. ({\bf C}) The single-sided amplitude of the Fourier spectrum calculated using the data of A, as a function of the effective detuning $\delta_\mathrm{eff}$.}
\end{figure}

\section*{Atom-Cavity Model}
We only consider the pump and cavity directions. The full atom-cavity system can be modeled using the many-body Hamiltonian with four terms describing the cavity, the atoms, and the atom-cavity interactions, given by
\begin{equation}
\hat{H} =\hat{H}_\mathrm{c}+\hat{H}_\mathrm{a}+\hat{H}_\mathrm{aa}+\hat{H}_\mathrm{ac}\, ,
\end{equation}
where the cavity contribution is $\hat{H}_\mathrm{c}=-\hbar \delta_{c}~\hat{a}^\dagger\hat{a}$ and the detuning between the pump and cavity frequencies is $\delta_\mathrm{c}<0$. The cavity mode annihilation and creation operator are denoted by $\hat{a}$ and $\hat{a}^\dagger$. The atomic part is described by
\begin{equation}
\hat{H}_\mathrm{a} = \int dy dz~ \hat{\Psi}^\dagger(y,z) \left(-\frac{\hbar^2}{2m} \nabla^2+\mathrm{V}_\mathrm{ext}(y,z)\right) \hat{\Psi}(y,z)
\end{equation}
where the external potential due to the standing wave created by the pump beam is $\mathrm{V}_\mathrm{ext}(y,z)=\epsilon_\mathrm{f} \cos^2(ky)$ with the potential strength parameter $\epsilon_\mathrm{f}$ and $m$ the mass of an atom.
The short-range collisional interaction between the atoms can be captured via
\begin{equation}
\hat{H}_\mathrm{aa}= U_\mathrm{a} \int dy dz~ \hat{\Psi}^\dagger(y,z) \hat{\Psi}^\dagger(y,z) \hat{\Psi}(y,z)\hat{\Psi}(y,z),
\end{equation}
where $U_\mathrm{a}=\sqrt{2 \pi}a_\mathrm{s}\hbar/m l_\mathrm{x}$ is the effective 2D interaction strength with $a_\mathrm{s}$ the s-wave scattering length and $l_\mathrm{x}$ the harmonic oscillator length in the x direction. 
The atom-cavity interaction part is described by
\begin{equation}
\hat{H}_\mathrm{ac}= \int dy dz~ \hat{\Psi}^\dagger(y,z)  \left(\hbar U_0\cos^2(kz)\hat{a}^\dagger\hat{a}+\hbar \sqrt{\hbar\epsilon_\mathrm{f}U_0}\cos(ky)\cos(kz)\left[\hat{a}^\dagger+\hat{a} \right] \right) \hat{\Psi}(y,z).
\end{equation}
The light shift per intracavity photon is denoted by $U_0>0$. For our numerical simulations of the dynamics, we use the semiclassical method based on the truncated Wigner approximation (TWA) \cite{POLKOVNIKOV, Cosme2020}. TWA approximates the quantum dynamics by solving the equations of motions over an ensemble of initial states, which are sampled from the initial Wigner distribution. This methods allows us to incorporate the leading order quantum corrections to the meanfield solution. The c number equation for the light field is
\begin{equation}
i \frac{\partial \alpha}{\partial t}= \frac{1}{\hbar}\frac{\partial H}{\partial \alpha^*}-i\kappa \alpha+i \xi=(-\delta_{c} +U_0 \mathcal{B}-i \kappa+i \xi)+\sqrt{\hbar\epsilon_\mathrm{f}U_0} \Phi, 
\end{equation}
where we have defined the bunching parameter $\mathcal{B} = \int dy dz~\cos^2(kz)|\psi(y,z)|^2$ and the density wave order parameter that corresponds to a checkerboard ordering $\Phi= \int dy dz~\cos(ky)\cos(kz)|\psi(y,z)|^2$. We further included a decay term proportional to $\kappa$ in the cavity mode dynamics and the resulting stochastic noise term $\xi(t)$, which is defined via $\langle\xi^*(t)\xi(t')\rangle=\kappa \delta(t-t')$. We obtain the atom-field equations via
\begin{equation}
i \frac{\partial \psi(y,z)}{\partial t}= \frac{1}{\hbar}\frac{\partial H}{\partial \psi^*(y,z)} = \left(-\frac{\hbar}{2m}\nabla^2+V_\mathrm{dip}(y,z)+2U_\mathrm{a} |\psi(y,z)|^2 \right)\psi(y,z)
\end{equation}
with
\begin{equation}
V_\mathrm{dip}(y,z)=\hbar \left(U_0|\alpha|^2\cos^2(kz) +\epsilon_\mathrm{f} \omega_\mathrm{rec} \cos^2(ky)+\sqrt{\hbar\epsilon_\mathrm{f}U_0}\left[\alpha +\alpha^* \right]\cos(ky)\cos(kz)\right).
\end{equation}
For the simulations we use the same set of parameters as in the experiment.

\section*{Breaking of continuous time translation symmetry}
To gain further insights into the continuous time translation symmetry breaking, we consider three different possibilities for including quantum noise in our theory. First, we sample over the full initial Wigner distribution and also include the corresponding stochastic noise $\xi$ corresponding to the cavity-field decay rate $\kappa$. Secondly, we include only the sampling of the Wigner distribution of the initial state and ignore the stochastic noise in time due to the fluctuation-dissipation term in the cavity field. Third, we fix the initial state and include stochastic noise in the cavity mode. For each case, we consider $10^3$ trajectories but for clearer presentation we only show the first 500 trajectories in Fig.~S2(A-C). To obtain Fig.~S2(A-C), we use $\delta_\mathrm{eff}=-2\pi \times 10.4~\mathrm{kHz}$ and linearly ramp up the pump strength to its final value $\epsilon_\mathrm{f}/\omega_\mathrm{rec}=0.85$ within $10~\mathrm{ms}$. We compute the fast Fourier transformation between $t_\mathrm{start}=15~\mathrm{ms}$ and $t_\mathrm{final}=65~\mathrm{ms}$. We record every $0.00125~\mathrm{ms}$ and thus, our frequency resolution is limited by $\Delta_\mathrm{FFT}=20~\mathrm{Hz}$. In Fig.~S2A and S2B the limit cycle frequency varies $\pm 150~\mathrm{Hz}$. For the data set in Fig.~S2C, the frequency is fixed. To minimize the fluctuations in the FFT signal due to the offset at $\omega=0$ we normalize each trajectory by the maximum of the FFT. For better accessibility, after obtaining the data from all trajectories we average over the mean value of all points. The TWA results in Fig.~S2A nicely show that all phases between $0$ and $2 \pi$ are realized. The same holds true in Fig.~S2B and Fig.~S2C. This suggests that the initial quantum noise and stochastic noise from the leaky cavity are sufficient to exhibit the breaking of continuous time translation symmetry.

{
\begin{figure}[h!]
	\centering
	\includegraphics[width= 0.85\columnwidth]{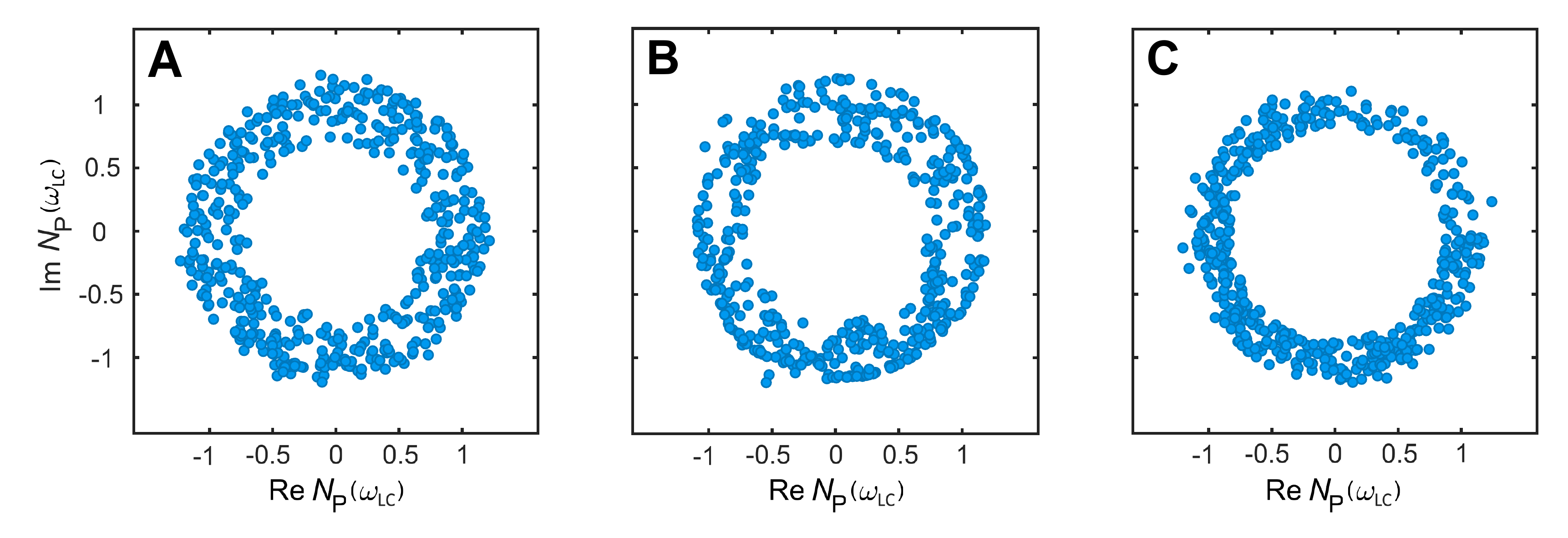}
	\label{sfig:2} 
\caption{{\bf  Distribution of the time phase in the limit cycle phase.} TWA simulations including ({\bf A})  both initial quantum noise and stochastic noise, ({\bf B})  only initial quantum noise, and ({\bf C})  only stochastic noise. We use $\delta_\mathrm{eff}=-2\pi \times 10.4~\mathrm{kHz}$ and $\epsilon_\mathrm{f}/\omega_\mathrm{rec}=0.85$.}
\end{figure}

\section*{Atom dynamics during one limit cycle}
We present the dynamics of the light field and the relevant density wave order parameters for a single exemplary trajectory in the limit cycle phase. We use $\delta_\mathrm{eff}=-2\pi \times 10.4~\mathrm{kHz}$ and a final pump strength of $\epsilon_\mathrm{f}/\omega_\mathrm{rec}=0.85$. We ramp up the pump intensity within $10~\mathrm{ms}$ and present in Fig.~S3C the limit cycle dynamics after $20~\mathrm{ms}$. We find that the only non-zero order parameters are those associated to the chequerboard density wave, $\Phi=\langle \cos(ky)\cos(kz)\rangle$, and to the density waves related to the cavity and pump bunching parameters, $\mathcal{B}=\langle \cos(kz)^2\rangle$ and $\mathcal{P}=\langle \cos(ky)^2\rangle$, respectively. Fig.~S3C shows the dynamics of the light field and the three order parameters. The oscillations in the dynamics of the atomic field density wave order parameter lags behind those in the cavity field occupation. In Fig.~S3(A-B) and Fig.~S3(C-D), the density of the atomic-field is presented. The atoms slosh back and forth from a checkerboard pattern to the minima of the light field intensity.

\begin{figure}[h!]
	\centering
	\includegraphics[width= 0.5\columnwidth]{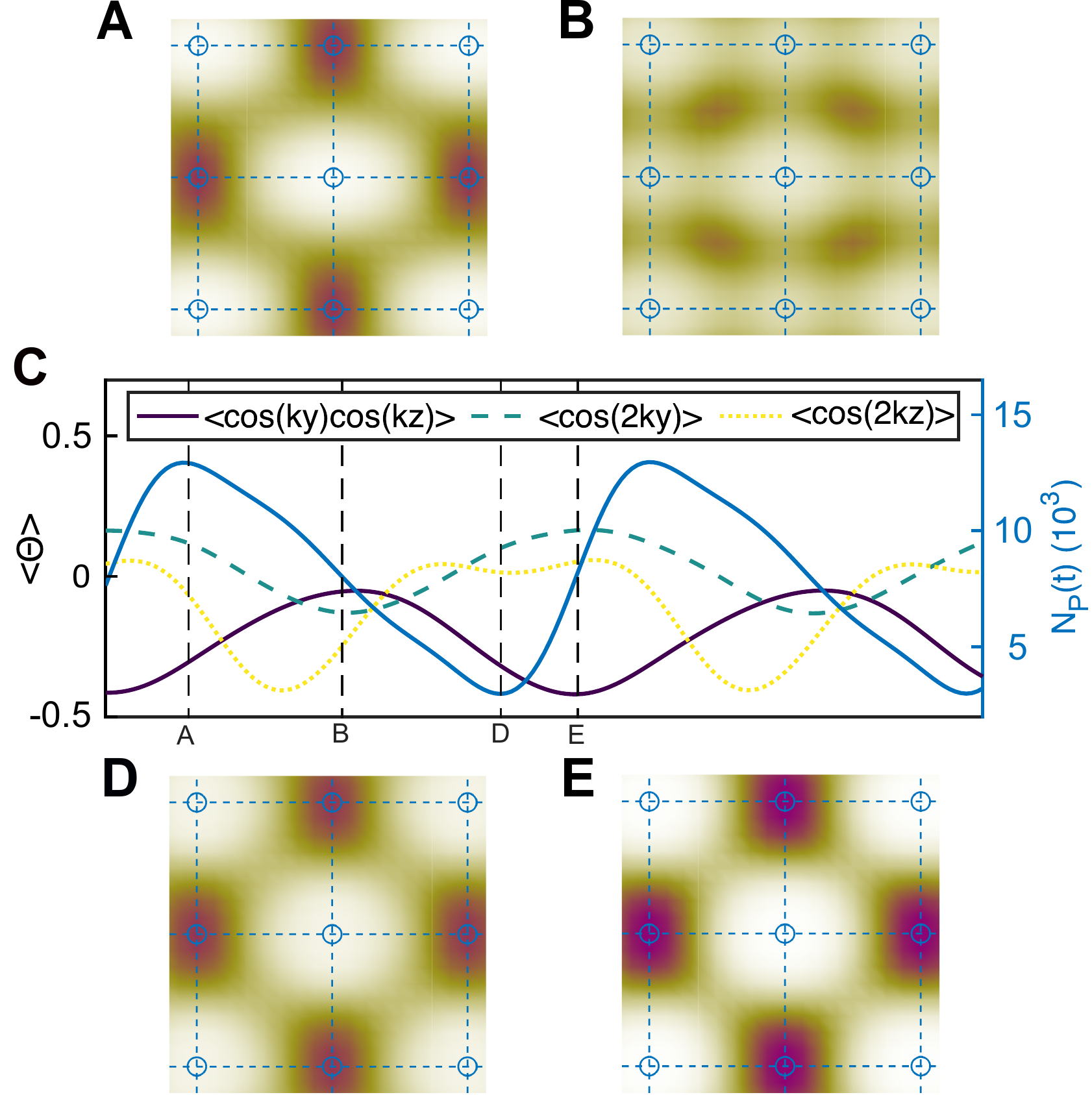}
	\label{sfig:3}
\caption{{\bf  Numerical results for the limit cycle dynamics.} ({\bf A-B}) and ({\bf D-E}) Atomic density distributions for different times during the limit cycle. The gray dashed lines in C indicate the times for which the density distributions are calculated. 
Horizontal and vertical dashed blue lines mark the extrema of $\cos(ky)$ and $\cos(kz)$, respectively and solid blue circles denote the extrema of the product $\cos(ky)\cos(kz)$, which determines the chequerboard density wave order parameter $\Phi$. {\bf C} Dynamics of the three relevant order parameters and the cavity mode occupation. The vertical dashed lines denote the times when (A-B) and (D-E) are taken. We use $\delta_\mathrm{eff}=-2\pi \times 10.4~\mathrm{kHz}$ and $\epsilon_\mathrm{f}/\omega_\mathrm{rec}=0.85$.}
\end{figure}

\section*{Stability against short-range interactions and atom losses}
We present the stability of the limit cycles against short-range interactions and phenomenological atom losses.
We measure the interaction strengths via the mean-field collisional interaction energy \cite{Kessler2021}
\begin{equation}
E_\mathrm{a}=\frac{U_\mathrm{a}}{N_\mathrm{a}} \int dy dz~ |\psi_0(y,z)|^4~
\end{equation}
with the wavefunction of the homogeneous BEC $\psi_0$. We further add a phenomenological atom loss term to our equations of motion of the form of
\begin{equation}
\frac{d N_\mathrm{a}}{dt}=-2 \gamma N_\mathrm{a} 
\end{equation}
to capture the atom losses in the experiment. To quantify the temporal long-range order we compute the two-point temporal correlation function 
\begin{equation}
C(t)=\mathrm{Re}\left(\frac{\langle \hat{a}^\dagger(t) a(t_0) \rangle }{\langle \hat{a}^\dagger(t_0) a(t_0) \rangle} \right).
\end{equation}
The time $t_0$ is defined as the time of the first maximum of the limit cycle oscillations after the transition into the superradiant phase.\\ 
We present the dynamics of the photon number $N_\mathrm{P}$ and the nonequal time correlation $C$ in Fig.~S4 for different collisional interaction strengths $E_\mathrm{a}$ and atom loss rates $\gamma$. We observe that short-range interactions do not destroy the temporal long range order for weaker collisional interaction energies $E_\mathrm{a}=0.1~E_\mathrm{rec}$ to strong interactions of $E_\mathrm{a}=0.2~E_\mathrm{rec}$. However, the combination of strong short-range interactions $E_\mathrm{a}=0.2~E_\mathrm{rec}$ and atom losses of $\gamma = 40~s^{-1}$ lead to a decay of the temporal order similar as observed in the experiment. The loss rate is chosen such that it models the observed atom decay rate in the experiment. We conclude that the main limitation of the limit cycle lifetime stems from atom losses in the experimental set up.

\begin{figure}[h!]
	\centering
	\includegraphics[width= 0.6\columnwidth]{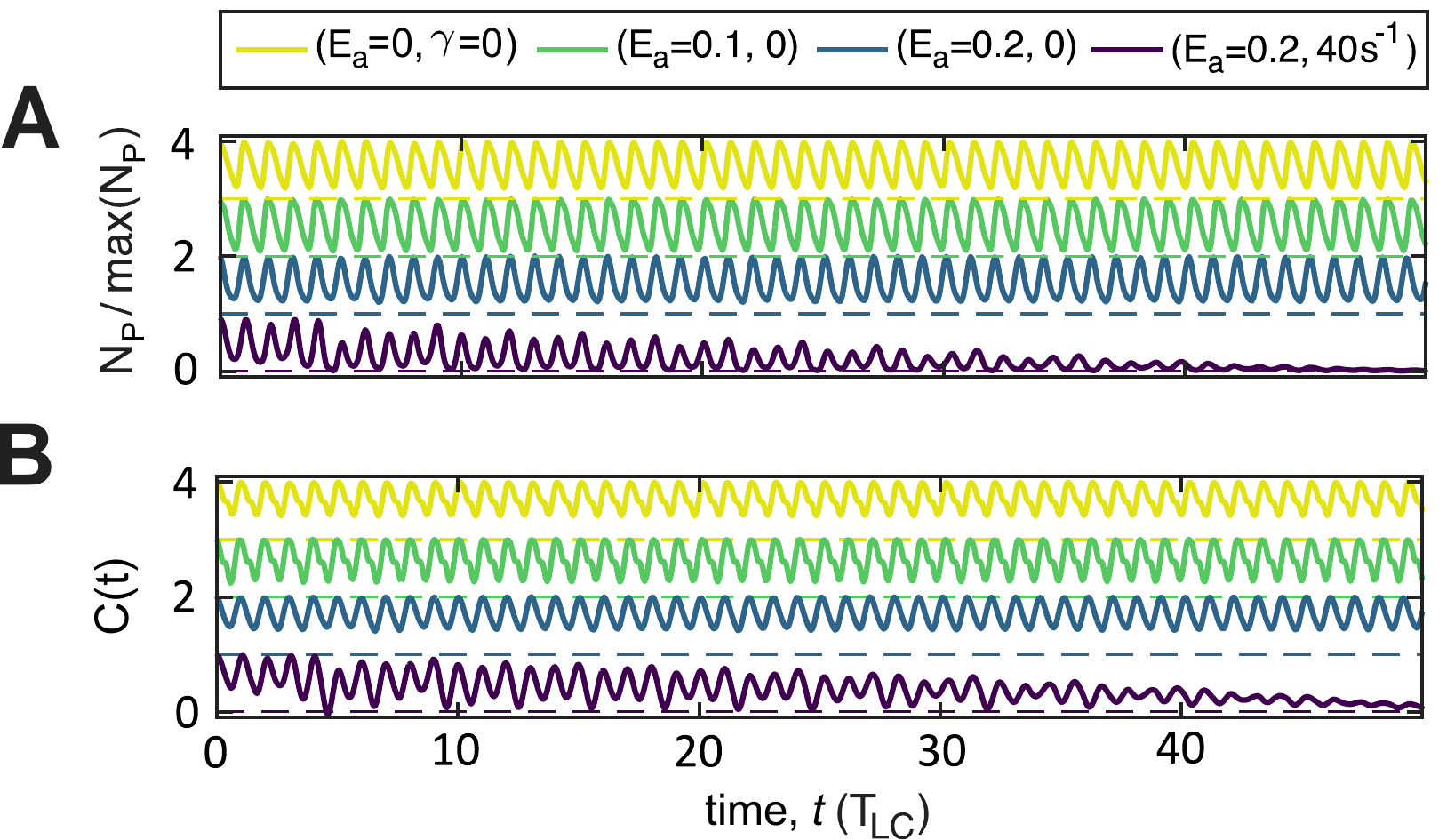}
	\label{sfig:4} 

\caption{{\bf  Numerical results on short-range interactions and atom losses.} ({\bf A}) Numerical results on the intracavity photon number $N_\mathrm{P}$ and ({\bf B}) the corresponding nonequal time correlation $C$ for different contact interaction energies $E_\mathrm{a}$ and atom losses $\gamma$. For better readability, we include an offset of $1,2,3$ for the blue, green and yellow trace indicated by the dashed lines. We fix $\delta_\mathrm{eff}=-2\pi \times 10.4~\mathrm{kHz}$.}
\end{figure}

\section*{Stability with respect to pump-atom detuning}
The pump-atom detuning is in our system parametrized by the single photon light shift $U_0$. For all the measurements presented in the main text $U_0 = 2\pi \times 1.3$~Hz is kept constant. To demonstrate robustness with respect to the pump-atom detuning, and hence with respect to $U_0$, we present in Fig.~S5 measurements of self-organization phase diagram for $U_0 = 2\pi \times 1.9$~Hz. The limit cycles are indicated by a peak in the Fourier spectrum of the intracavity photon number (Fig.~S5C), which can be found for small negative effective pump-cavity detuning $\delta_\mathrm{eff}/2\pi$, between $-10$ and $-20\,\mathrm{kHz}$. This measurement is only an example and we experimentally observe stable limit cycles for different values of $U_0$.

\begin{figure}[h!]
	\centering
	\includegraphics[width= 1\columnwidth]{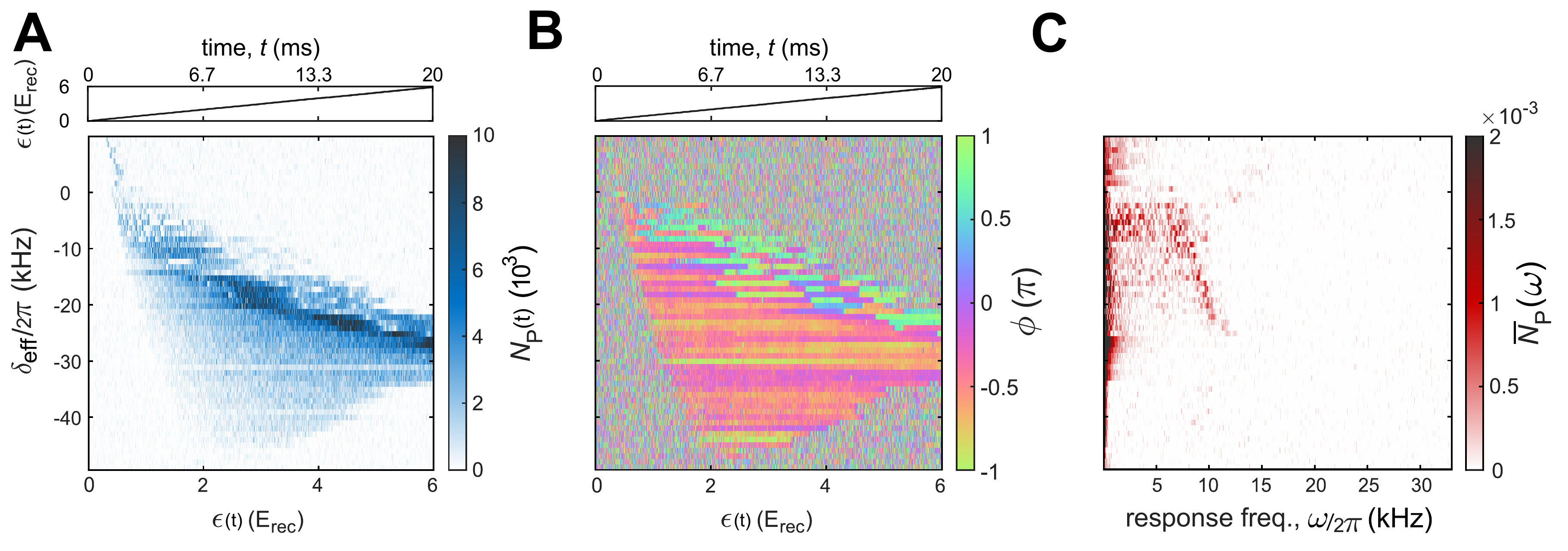}
	\label{sfig:5} 
	\caption{{\bf  Phase diagrams for another pump-atom detuning.}   ({\bf A})  Top panel: Pump strength protocol. Bottom panel: The corresponding intracavity photon number $N_\mathrm{P}$, as a function of the effective detuning $\delta_\mathrm{eff}$ and pump strength $\epsilon$ at a pump wavelength of $\lambda_\mathrm{P} = 793.76\,$nm. The corresponding light shift per photon is $U_0 = 2\pi \times 1.9~\mathrm{Hz}$. ({\bf B}) Top panel: Pump strength protocol. Bottom panel: The phase difference between the pump and intracavity field $\phi$, as a function of the effective detuning $\delta_\mathrm{eff}$ and pump strength $\epsilon$. ({\bf C}) The single-sided amplitude of the Fourier spectrum calculated using the data of A, as a function of the effective detuning $\delta_\mathrm{eff}$. Red region around $8-10\mathrm{kHz}$ at small negative $\delta_\mathrm{eff}$ indicate a region where limit cycle can be found.}
\end{figure}

\section*{Finite-size effects}
We investigate the effects of a finite particle number on the stability of the time crystal. To this end, we compare the mean-field results, which simulate the thermodynamic limit, and the results of single TWA trajectories, which include stochastic noise associated to cavity loss, for different particle numbers. Owing to the cavity-induced all-to-all coupling between the atoms, the thermodynamic limit is expected to be captured by our mean-field theory. We vary the particle number while keeping $N U_0$ fixed. We obtain the peaks in the dynamics of the intracavity photon number, $\tilde{N}_\mathrm{P}$, to highlight the change in the oscillation amplitude of the limit cycle phase for varying particle number. In Fig.~S6A, we show the time evolution of $\tilde{N}_\mathrm{P}/N_\mathrm{a}$  for some exemplary particle numbers using TWA and the MF result corresponding to the thermodynamic limit. It can be seen that as the particle number is increased, the results approach the MF prediction. This means that the temporal dynamics becomes more regular as we increase the particle number $N_\mathrm{a}$ towards the thermodynamic limit. To further illustrate this point, we calculate the relative crystalline fraction $\Xi' \equiv \sum_{\omega \in \delta_\mathrm{LC}} N_\mathrm{P}(\omega)/\sum_{\omega \in \Delta_\mathrm{LC}} N_\mathrm{P}(\omega)$. We rescaled the relative crystalline fraction for varying $N_\mathrm{a}$ by the value in the thermodynamic limit, i.e., the $\Xi' $ in our mean-field prediction is set to 1 as indicated by the gray dashed line in Fig.~S6B. The blue cross marks the typical particle number in our experiment. We find that as $N_\mathrm{a}$ is increased, the crystalline fraction approaches the mean-field prediction. This can be understood from the fact that the initial quantum noise and stochastic noise scales with $1/N$ in TWA, meaning that as expected for $N_\mathrm{a} \rightarrow \infty$, we recover the thermodynamic limit.

\newpage

\begin{figure}[h!]
	\centering
	\includegraphics[width= 0.8\columnwidth]{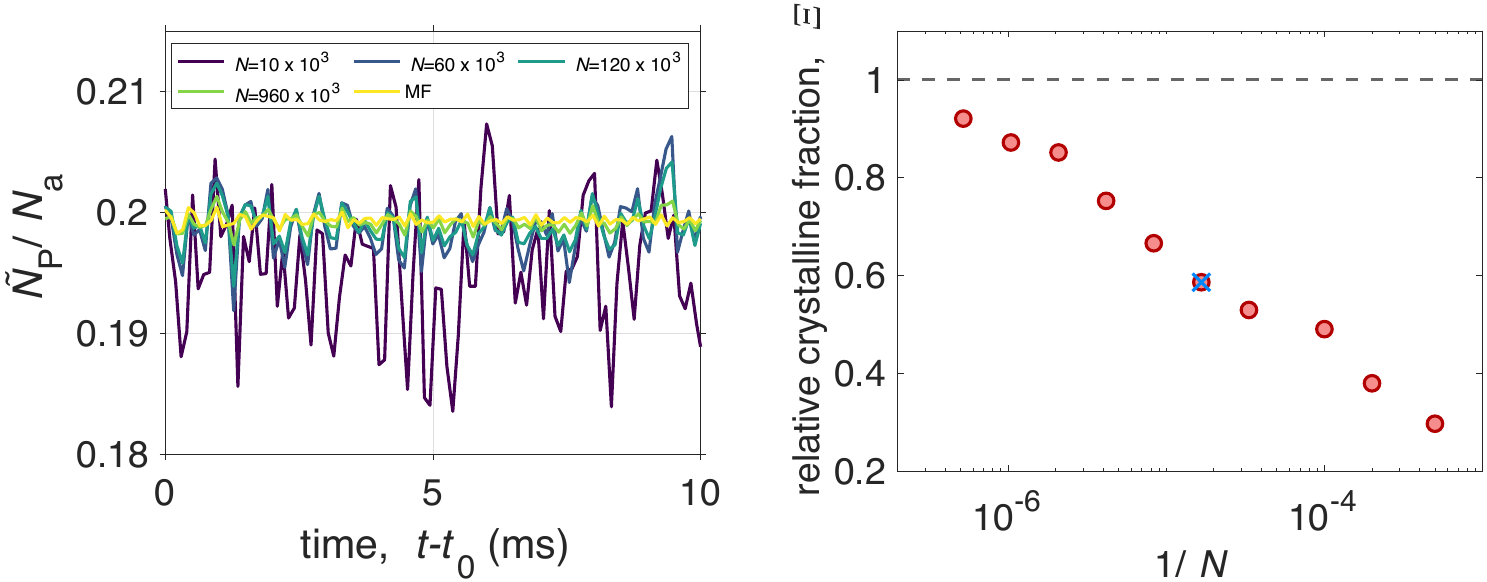}
	\label{sfig:6} 

\caption{{\bf Numerical results on the stability for different particle numbers} ({\bf A}) Results of a single TWA trajectory for the peak height of the intracavity photon number in the limit cycle phase for different particle numbers.  ({\bf B}) The relative crystalline fraction for varying particle numbers obtained from single  TWA trajectories, which include stochastic noise from the cavity losses. The gray horizontal dashed line represents the mean-field crystalline fraction, which we set to 1 as a benchmark for finite $N$. The blue cross indicates the particle number, in which the experiment operates.  We fix $\delta_\mathrm{eff}=-2\pi \times 10.4~\mathrm{kHz}$ and $N U_0=2\pi \times 60~\mathrm{kHz} = \mathrm{const.}$}
\end{figure}

\section*{Stability against temporal perturbations}
The stability of the limit cycle phase against temporal noise can be also explored using our theoretical model. We focus on the mean-field regime to show that the limit cycle phases in the thermodynamic limit exhibit the robustness expected of a continuous time crystal. We add a Gaussian white noise onto the pump signal, which is band-limited to $0.025~\mathrm{GHz}$. This is set by the integration step size of our stochastic differential equation solver. Note that the noise in the experiment is band-limited to $50~\mathrm{kHz}$. Examples of the noisy pump signal are shown in Fig.~S7A. 
The noise strength is quantified by a parameter similar to the one in the experiment, $n \equiv { \sum_{\omega} |\mathcal{A}_\mathrm{noisy}(\omega)| }/{\sum_{\omega} |\mathcal{A}_\mathrm{clean}(\omega)|}  -1$, where $\mathcal{A}$ is the Fourier spectrum of the pump signal. In Fig.~S7B, we show the peaks in the dynamics of the intracavity photon number, $\tilde{N}_\mathrm{P}(t)$, for various noise strengths. We find that increasing the temporal noise strength leads to more irregular oscillations in the limit cycle phase. To further quantify this behaviour, we again obtain the relative crystalline fraction as defined in the previous section. The dependence of the relative crystalline on temporal noise strength $n$ is shown in Fig.~S7C. We observe that for small noise strength, the crystalline fraction appears unchanged. The time crystal starts to melt for stronger noise strengths as expected. These numerical results qualitatively agree with the experimental results shown in Fig.~3E and they suggest the robustness of the limit cycle phase in the thermodynamic limit against temporal perturbation.

\begin{figure}[h!]
	\centering
	\includegraphics[width= 0.8\columnwidth]{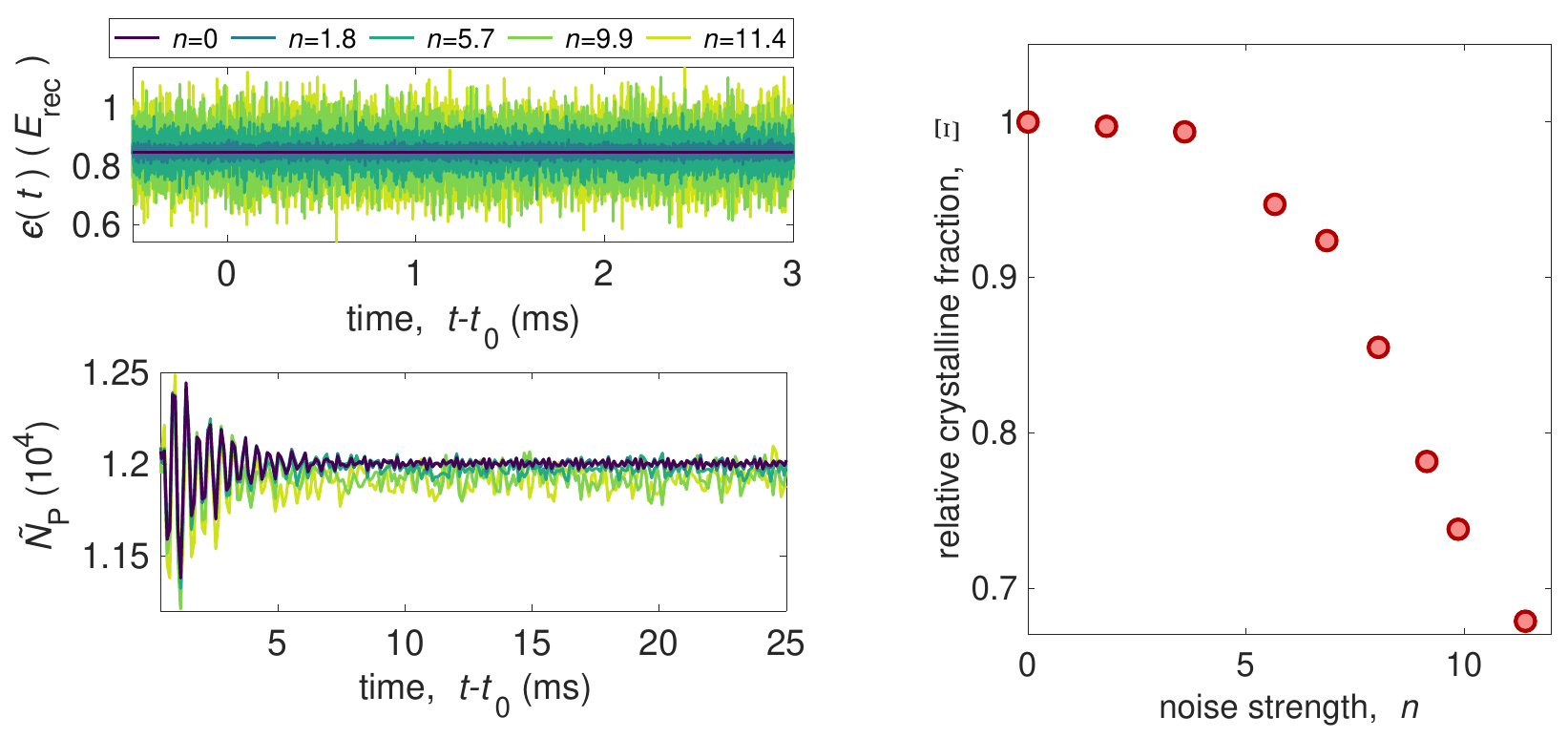}
	\label{sfig:7} 
\caption{{\bf  Numerical results on the stability against temporal noise} ({\bf A}) Time dependence of the pump strength $\epsilon$ for different noise strengths and ({\bf B}) the corresponding mean-field results for the dynamics of peak height of the intracavity photon number in the limit cycle phase. ({\bf C}) The relative crystalline fraction for different noise strength.}
\end{figure}

\section*{Route to chaos}
Our system exhibits a route to chaos, which we have investigated theoretically in a previous study. The full dynamical phase diagram including the chaotic regime can be found in Ref.~\cite{Kessler2019}. We find that the limit cycle phase becomes unstable towards chaotic dynamics for large pump strengths. Due to the limited lifetime of the BEC in our experimental setup, it is difficult to experimentally identify such a chaotic phase, which manifests in its characteristic long-time dynamics.